# Coherently excited superresolution using intensity product of phase-controlled quantum erasers via polarization-basis projection measurements

Byoung S. Ham[1,2]
[1]School of Electrical Engineering and Computer Science, Gwangju Institute of Science and Technology, 123 Chumdangwagi-ro, Buk-gu, Gwangju 61005, South Korea
[2]Qu-Lidar, 123 Chumdangwagi-ro, Buk-gu, Gwangju 61005, South Korea
(February 26, 2024; bham@gist.ac.kr)

**Abstract**
Recently, the delayed-choice quantum eraser has been applied for coherently excited superresolution using phase-controlled projection measurements of laser light to overcome the diffraction limit in classical physics as well as to solve the limited photon number of the N00N state in quantum physics. Unlike other methods of phase-controlled superresolution in a noninterferometric system, the proposed method is for the intensity products between phase-controlled quantum erasers, resulting in superresolution compatible with the most conventional sensing metrologies. Here, a general scheme of the phase-controlled quantum eraser-based superresolution is proposed and its general solution is derived for an arbitrary $N^{th}$-order intensity correlation, where the superresolution shows the photonic de Broglie wave-like quantum feature. Furthermore, phase quantization of the superresolution is discussed to better understand quantum mechanics.

**Introduction**
Quantum entanglement is between two or more individual particles, where a fixed phase relation between paired photons does not violate quantum mechanics [1]. A typical method of entangled photon pair generation is to use a spontaneous parametric down-conversion (SPDC) process [2], where the phase-matching condition among the pump and two sibling photons is critical [1,2,3,4]. Unlike a single photon, thus, the fixed phase between entangled photons is straightforward for the wave nature of quantum mechanics [1-7]. Such an understanding of the wave nature-based quantum correlation has emerged to revisit the Hong-Ou-Mandel (HOM) effect [5,6], Franson-type nonlocal correlation [7], and delayed-choice quantum eraser [8,9]. Experimental demonstration of the fixed phases relation has been conducted in trapped ions for a $\pi/2$ phase difference [10]. A complete coherence solution of the HOM effect for the $\pi/2$ phase relation has also been presented [5,6]. Most recently, the same phase relation has been applied to the superresolution whose fundamental physics is in the nonlocal correlation [11].

The wave-particle duality originates in quantum superposition, where these two natures are mutually exclusive [12-14]. In a single photon's self-interference [15], thus, the quantum superposition is between orthonormal bases of the single photon [16-19]. With the wave nature, the delayed-choice quantum eraser [17] has been newly interpreted [20], as an ad-hoc quantum superposition of orthonormal bases of a single photon through a dynamic window of a polarizer for the basis-projection measurement [21]. Due to the exclusive nature between the phase (wave) and photon number (particle), thus, the interpretation of the quantum eraser represents a deterministic quantum feature, where no difference exists between single photon [8] and continuous wave (cw) light [9]. Similarly, phase-controlled superresolution [22] has been experimentally demonstrated using a single photon and cw light for the same quantum feature of photonic de Broglie waves (PBWs) [23-29].

Here, a universal scheme of the phase-controlled superresolution is proposed for a cw laser in a Michelson interferometer. In quantum sensing metrology, the superresolution overcoming the shot-noise limit (SNL) has been experimentally demonstrated using higher-order entangled photon pairs such as N00N states [24-30] or squeezed state [31] to satisfy the Heisenberg limit (HL) [21-33]. The N00N state-based superresolution is known as PBWs [23-29]. Such a PBW-like superresolution effect has also been observed using phase-controlled coherent photons in a noninterferometric system [32,33] via projection measurements [21]. On the contrary, the proposed superresolution is for the intensity product of phase-controlled quantum erasers using cw laser. Here, a universal scheme of an arbitrary $N^{th}$-order superresolution is proposed, and its general solution is coherently derived for the intensity product of quantum erasers via projection measurements. Moreover, the superresolution is compared with PBW-like quantum features and discussed for phase quantization of the ordered intensity products.

**Result**

*A. Phase-controlled projection measurement of quantum erasers for superresolution*

Figure 1 shows a universal scheme of the classically excited superresolution based on phase-controlled quantum erasers. The superresolution scheme in Fig. 1 originates in the $N^{th}$-order intensity correlations between phase-controlled quantum erasers, resulting in the PBW-like quantum feature, as shown in Fig. 2. Compared to N=4 case [12], the Inset of Fig. 1 shows an arbitrary $N^{th}$-order superresolution scheme, where the first eight quantum erasers for N=8 are visualized with dotted blocks to explain the phase control of the quantum erasers. For the quantum eraser, both single photon [9] and cw laser light [10] were experimentally demonstrated in a Mach-Zehnder interferometer (MZI) for the polarization-basis projection onto a polarizer P. The MZI physics of coherence optics [34] shows the same feature in both a single photon [15] and cw light due to the limited Sorkin parameter, as discussed for the Born rule tests [35]. Quantum mechanically, the deterministic feature of the MZI system is due to the double unitary transformation of a 50/50 nonpolarizing beam splitter (BS) [1].

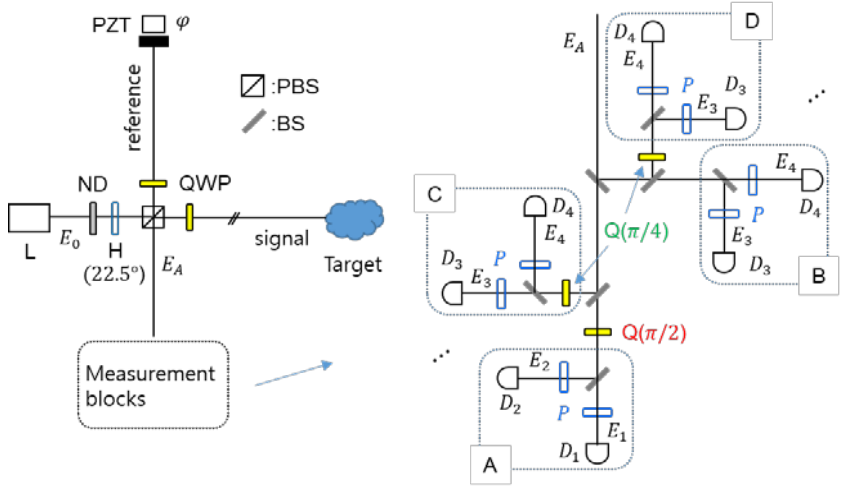

**Fig. 1. Schematic of a universal super-resolution based on phase-controlled quantum erasers.** L: laser, ND: neutral density filter, H: half-wave plate, PBS: polarizing beam splitter, PZT: piezo-electric transducer, QWP: quarter-wave plate, P: polarizer, D: single photon (or photo-) detector, All rotation angles of Ps are independent at $\theta = 45°$.

The rotation angle of quarter-wave plates (QWPs) inserted in blocks C and D is half that of block A. Thus, the QWP-resulting phase gain to the vertical component of the light is also half the block A [34]. This polarization-basis-dependent phase gain of the light directly affects the quantum eraser, because the role of the polarizer P is to project orthogonal polarization bases onto the common axis $\hat{p}$ (see Eqs. (2)-(8)) [19]. The random path length to the quantum eraser does not influence the intensity correlations due to the unaffected global phase by the Born rule, where intensity (measurement) is the absolute square of the amplitude [13,14]. By controlling the relative phase between blocks, the first-order intensity products of quantum erasers can be appropriately shifted for the superresolution (see Fig. 2). For the proposed universal scheme of practically infinite number of phase-controlled quantum erasers in Fig. 1, a general coherence solution of an arbitrary $N^{th}$-order superresolution is coherently derived (see Eq. (24) and Fig. 3). And then, the general solution is compared with PBWs based on N00N states for the discussion of phase quantization of the $N^{th}$-order intensity correlation.

Such phase quantization has already been separately discussed for coherence de Broglie waves (CBWs) in a coupled MZI system [36,37]. Unlike CBWs resulting from MZI superposition, the present phase quantization of superresolution is for the intensity product of phase-controlled quantum erasers. On the contrary to energy quantization in the particle nature of quantum mechanics [1], the phase quantization is for the wave nature.

### B. Analysis 1: PBW-like superresolution

A coherence approach based on the wave nature of a photon is adopted to analyze Fig. 1 differently from the quantum approach based on quantum operators [1,23-30]. The novel feature of the present method is to use common intensity products of cw lights via polarization-basis projection of the phase controlled quantum erasers. Thus, there is no need for single-photon coincidence detection. Instead, the intensity product is enough for a single shot measurement, as in nonlinear optics. Technically, the condition $M \geq N$ is required, and M is the number of the divided quantum erasers, where M should be far less than the total number of photons of the input laser for the classical approach. In the use of a time-bin scheme, a pulse laser is more appropriate, as shown for quantum key distribution [38].

The amplitude of the output field of the Michelson interferometer in Fig. 1 is represented as:

$$\boldsymbol{E}_A = \frac{iE_0}{\sqrt{2}}(\hat{H} + \hat{V}e^{i\varphi}), \qquad (1)$$

where $E_0$ is the amplitude of cw light from L. $\hat{H}$ and $\hat{V}$ are unit vectors of horizontal and vertical polarization bases of the light, respectively. Unlike correlated (entangled) single-photon number N entering the interferometer for the $N^{th}$-order intensity correlation [28], the cw light (or N single photons) is simply divided into M ports for the classical (coincidence) detection of the quantum erasers [39]. Due to the orthogonal bases, Eq. (1) results in no fringe: $\langle I_A \rangle = I_0$.

By the rotated polarizers in Fig. 1, whose rotation angle θ is from the horizontal axis, Eq. (1) is modified for the quantum eraser:

$$\boldsymbol{E}_{A1} = \frac{iE_0}{\sqrt{2}\sqrt{8}}(cos\theta e^{i\varphi} + isin\theta)\hat{p}, \qquad (2)$$

$$\boldsymbol{E}_{A2} = \frac{iE_0}{\sqrt{2}\sqrt{8}}(-cos\theta e^{i\varphi} + isin\theta)\hat{p}, \qquad (3)$$

$$\boldsymbol{E}_{B1} = \frac{-iE_0}{\sqrt{2}\sqrt{8}}(cos\theta e^{i\varphi} + sin\theta)\hat{p}, \qquad (4)$$

$$\boldsymbol{E}_{B2} = \frac{-E_0}{\sqrt{2}\sqrt{8}}(-cos\theta e^{i\varphi} + sin\theta)\hat{p}, \qquad (5)$$

$$\boldsymbol{E}_{C1} = \frac{-E_0}{\sqrt{2}\sqrt{8}}(-cos\theta e^{i\varphi} + sin\theta)\hat{p}, \qquad (6)$$

$$\boldsymbol{E}_{C2} = \frac{-iE_0}{\sqrt{2}\sqrt{8}}(cos\theta e^{i\varphi} + sin\theta)\hat{p}, \qquad (7)$$

$$\boldsymbol{E}_{D1} = \frac{-E_0}{\sqrt{2}\sqrt{8}}(cos\theta e^{i\varphi} + sin\theta)\hat{p}, \qquad (8)$$

$$\boldsymbol{E}_{D2} = \frac{-E_0}{\sqrt{2}\sqrt{8}}(-cos\theta e^{i\varphi} + sin\theta)\hat{p}, \qquad (9)$$

where $\hat{p}$ is the axis of the polarizers, and $\sqrt{8}$ is due to the eight divisions (M=8) of $\boldsymbol{E}_A$ by the lossless BSs. In Eqs. (2)-(9), the projection onto the polarizer results in $\hat{H} \rightarrow cos\theta\hat{p}$ and $\hat{V} \rightarrow sin\theta\hat{p}$. By BS, the polarization direction of $\hat{H}$ is reversed, as shown in the mirror image [34].

Thus, the corresponding mean intensities are as follows for $\theta = 45°$ of all Ps:

$$\langle I_{A1}\rangle = \frac{I_0}{2M}\langle 1 + sin\varphi\rangle, \tag{10}$$

$$\langle I_{A2}\rangle = \frac{I_0}{2M}\langle 1 - sin\varphi\rangle, \tag{11}$$

$$\langle I_{B1}\rangle = \frac{I_0}{2M}\langle 1 + cos\varphi\rangle, \tag{12}$$

$$\langle I_{B2}\rangle = \frac{I_0}{2M}\langle 1 - cos\varphi\rangle, \tag{13}$$

$$\langle I_{C1}\rangle = \frac{I_0}{2M}\langle 1 - cos(\varphi - \xi_1)\rangle, \tag{14}$$

$$\langle I_{C2}\rangle = \frac{I_0}{2M}\langle 1 + cos(\varphi - \xi_1)\rangle, \tag{15}$$

$$\langle I_{D1}\rangle = \frac{I_0}{2M}\langle 1 + cos(\varphi - \xi_2)\rangle, \tag{16}$$

$$\langle I_{D2}\rangle = \frac{I_0}{2M}\langle 1 - cos(\varphi - \xi_2)\rangle, \tag{17}$$

where $\xi_1$ and $\xi_2$ are rotation angles of QWPs in blocks C and D, respectively. Thus, the quantum erasers are analytically confirmed for cw light, resulting in interference fringes. The quantum mystery of the cause-effect relation of the quantum eraser can be found in the ad-hoc polarization-basis superposition via the polarization projection onto the $\hat{p}$ axis of the polarizer. The price to pay for this quantum mystery is 50% photon loss [8,9].

The corresponding second-order (N=2) intensity correlations between the quantum erasers in each block is directly obtained from Eqs. (10)-(17):

$$\langle C^{(2)}_{A1A2}(0)\rangle = \frac{I_0^2}{2^2 M^2}\langle cos^2\varphi\rangle, \tag{18}$$

$$\langle C^{(2)}_{B1B2}(0)\rangle = \frac{I_0^2}{2^2 M^2}\langle sin^2\varphi\rangle, \tag{19}$$

$$\langle C^{(2)}_{C1C2}(0)\rangle = \frac{I_0^2}{2^2 M^2}\langle sin^2(\varphi - \xi_1)\rangle, \tag{20}$$

$$\langle C^{(2)}_{D1D2}(0)\rangle = \frac{I_0^2}{2^2 M^2}\langle sin^2(\varphi - \xi_2)\rangle. \tag{21}$$

Thus, the corresponding fourth-order (N=4) intensity correlations between any two blocks can be represented from Eqs. (18)-(21):

$$\langle C^{(4)}_{A1A2B1B2}(0)\rangle = \frac{I_0^4}{2^6 M^4}\langle sin^2 2\varphi\rangle, \tag{22}$$

$$\langle C^{(4)}_{C1C2D1D2}(0)\rangle = \frac{I_0^4}{2^6 M^4}\langle sin^2(\varphi - \xi_1)sin^2(\varphi - \xi_2)\rangle. \tag{23}$$

Finally, the eighth-order (N=8) intensity correlation for all quantum erasers in Fig. 1 is as follows:

$$\langle C^{(8)}_{A1A2B1B2C1C2D1D2}(0)\rangle = \frac{I_0^8}{2^{12} M^8}\langle sin^2 2\varphi sin^2(\varphi - \xi_1)sin^2(\varphi - \xi_2)\rangle. \tag{24}$$

*C. Analysis II: Numerical calculations of the superresolution*

Figure 2 shows numerical calculations of the N$^{th}$-order intensity correlations using Eqs. (2)-(9) for $\xi_1 = \xi_2 = \pi/4$ to demonstrate the super-resolution using phase-controlled coherent light of L in Fig. 1. The left panel of Fig. 2 is for the first-order (N=1) intensity correlations of all quantum erasers, satisfying the phase-spaced fringes among them. In each block, a $\pi$ phase shift is shown (see the *-mark curves), implying fringe doubling for the second-order intensity correlations.

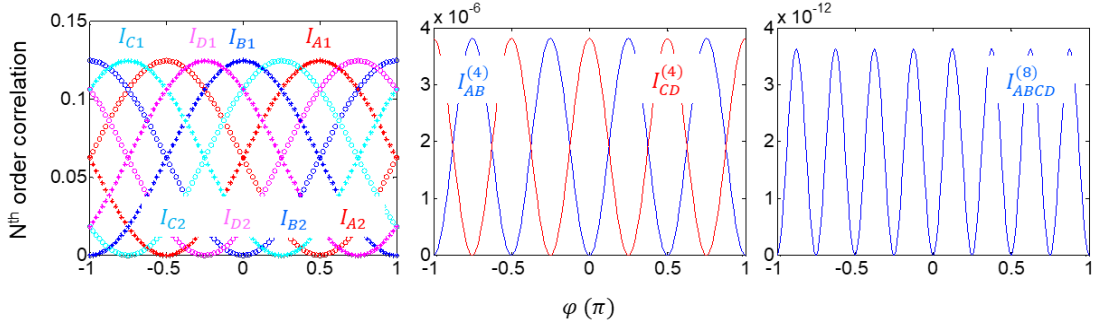

**Fig. 2. Numerical calculations of the N$^{th}$ order intensity correlations in Fig. 1.** (left) Individual first-order intensity correlation $I_j$ in A, B, C, and D blocks. Blue star (circle): $B_3$ ($B_4$) in B, Cyan star (circle): $C_3$ ($C_4$) in C, Red star (circle): $A_3$ ($A_4$) in A, Magenta star (circle): $D_3$ ($D_4$) in D. (middle) Fourth-order intensity correlation between (red) A and B, and (blue) C and D. (right) Eight-order intensity product between all quantum erasers. $\xi_1 = \xi_2 = \pi/4$ and $I_A^{(2)} = I_{A1} I_{A2}$.

The middle panel of Fig. 2 is for the fourth-order intensity correlations between blocks A and B (blue curve), and blocks C and D (red curve). Compared with the left panel, the number of fringes is quadrupled satisfying the Heisenberg limit of PBWs [28]. The right panel of Fig. 2 is for the eight-order intensity correlation between all blocks, where the fringes are eight times denser than that of each quantum eraser in the left panel. Thus, the phase resolution of the N$^{th}$-order intensity correlations satisfies the $\pi/N$ relation of the quantum sensing [28]. This phase resolution of the phase-controlled quantum erasers in Fig. 1 is the same as PBWs by N00N states [23-29]. Thus, Fig. 2 demonstrates the superresolution satisfied by PBWs in quantum sensing.

Figure 3 is for the details of numerical calculations for N=1~8 and N=80. The top panels of Fig. 3 are for odd and even Ns, where the fringe number linearly increases as N increases. Based on this N-proportional fringe number, the positions of the first fringes of N=1~8 move from $\pi/2$ for N=1 (black dot) to $\pi/16$ for N=8 (blue dot). As in PBWs, thus, the same interpretation of the N-times increased effective frequency compared to the input light is drown for these PBW-like N$^{th}$-order intensity correlations. Due to the post-measurements of intensity product, however, phase quantization is rather reasonable to the wave nature, as compared to the energy quantization in the particle nature of quantum mechanics. Again, the wave and particle natures are mutually exclusive.

For an arbitrary N, the j$^{th}$ block with $\xi_j$ QWP can be assigned for a universal scheme of the superresolution. For the expandable finite block series with $\xi_j$-phase-controlled quantum erasers in Fig. 1, the generalized solution of the N$^{th}$-order intensity correlation is as follows:

$$\langle C^{(N)}_{P_1 P_2 \ldots P_j \ldots P_N}(0) \rangle = \frac{I_0^N}{2^N M^N} \langle \prod_{j=1}^{N} sin^2\varphi \, sin^2(\varphi - \xi_j) \rangle, \quad (25)$$

where $\xi_j = \pi/2j$ and $sin^2\varphi$ term is for the reference of B block.

The bottom panels of Fig. 3 are for comparison purposes between N=8 and N=80, where the resulting ten times increased fringe numbers indicates ten times enhanced phase resolution. Thus, the pure coherence solution of the PBW-like quantum feature satisfying Heisenberg limit is numerically confirmed for the generalized solution in Eq. (25). Here, the coincidence detection in the particle nature of quantum sensing with N00N states is equivalent to the coherence intensity-product measurement, where the coherence between quantum erasers is well provided within the coherence length of the laser L.

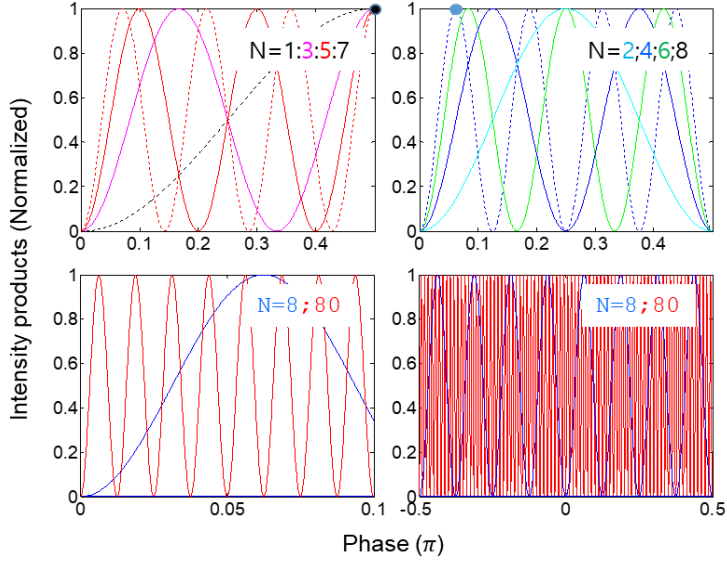

**Fig. 3. Numerical calculations for the normalized $N^{th}$-order intensity products between the phase-controlled quantum erasers in Fig. 1.** N is the number of quantum erasers to be used for intensity product measurements.

Figure 4 shows the respective phase-basis relation of the $N^{th}$-order intensity products for N=1, 2, 4, and 8 in Fig. 3. From the colored dots representing the first fringes of the ordered intensity products, the generalized phase basis of the $j^{th}$-order intensity product is derived for $\varphi_j = \pi/j$. This phase-basis relation applies to all js, where j=1,2,3… As shown, the $j^{th}$-order intensity correlation behaves as a j-times increased frequency of input light for the Michelson interferometer. This order-proportional effective frequency shows the PBW-like quantum feature of the N00N state in quantum metrology [23-29].

Based on the j-times repeated fringes in the $j^{th}$-order intensity product for $0 \leq \varphi \leq 2\pi$, the numerical results in Fig. 4 can be interpreted as of phase quantization of the intensity products through projection measurements of the quantum erasers. As shown in the PBW-like quantum features, these discrete eigenbases of the intensity products can also be compared to an N-coupled pendulum system [40], where the phase quantization in Fig. 4 can be classically understood [36,37]. Unlike the N-coupled pendulum system [40] or CBWs [36,37], any specific mode of $\varphi_j$ can be deterministically taken out by choosing a particular number of blocks for the intensity-product order N in Fig. 4. Like the energy quantization in the particle-nature-based quantum mechanics, thus, Fig. 4 is another viewpoint of the phase quantization in the wave nature.

From the universal scheme of the superresolution based on the phase-controlled quantum erasers in Fig. 1, a generalized solution of the $N^{th}$-order intensity correlation for Fig. 4 can also be intuitively obtained:

$$\langle C^{(N)}_{P_1 P_2 \ldots P_j \ldots P_{N/2}}(0) \rangle = \frac{I_0^N}{2^N M^N} \langle sin^2(N\varphi/2) \rangle, \tag{26}$$

where $P_j = Z_1 Z_2$ and $Z_k$ is the k$^{th}$ quantum eraser of the Z block: k=1,2. Here, the effective phase term $N\varphi$ in Eq. (26) represents the typical nonclassical feature of PBWs used for quantum sensing with N00N states [27,28]. The numerical simulations of Eq. (26) for N=1, 2, 4, and 8 perfectly match those in Fig. 4, respectively. Although the mathematical forms between Eqs. (25) and (26) are completely different, their quantum behaviors are the same as each other. Thus, Eq. (26) is equivalent to the superresolution in Eq. (25) [13,22], where the phase quantization is accomplished by ordered intensity products of the divided output fields of the typical interferometer. Unlike coincidence detection between entangled photons under the particle nature [23-29], the present coherence scheme with the wave nature is intrinsically deterministic within the coherence length of the input laser. Thus, the coincidence detection is replaced by the coherence product between independently controlled output fields. Such a projection measurement of the individually and independently controlled quantum erasers can also be applied for a time-bin scheme with a pulsed laser, where different time bins are completely ignored due to their incoherence feature [38].

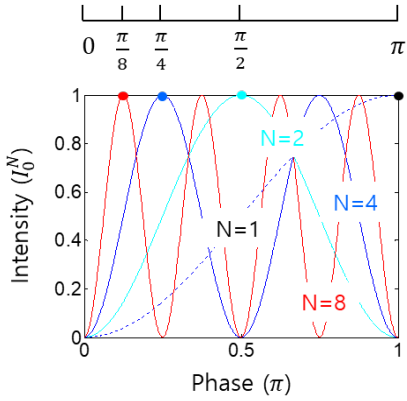

**Fig. 4. Phase quantization of the intensity products in Fig. 3.** N is the order of intensity product. Dotted: N=1, Cyan: N=2, Blue: N=4, Red: N=8. Intensity is normalized.

**Conclusion**

A universal scheme of superresolution was presented for the intensity product of the phase-controlled quantum erasers via polarization-basis projection measurements in a Michelson interferometer. The related general solution of the superresolution was coherently derived from the universal scheme for the intensity product. For the phase control of the quantum erasers, N different QWPs were assigned to the N-divided output fields whose rotation angles were set to be discretely decreased. As a result, all individual first-order intensity correlations of the N phase-controlled quantum erasers satisfied non-overlapped fringes in the phase domain, resulting in superresolution of the PBW-like quantum feature for their intensity products. Furthermore, the ordered intensity products were interpreted as phase quantization of the wave nature, as the energy quantization in the particle nature of quantum mechanics. Thus, the new concept of phase quantization was inherently deterministic due to the wave nature. Although such quantized phase modes were found in an N-coupled pendulum system, a deterministic choice of a particular eigenmode of the superresolution was possible by choosing a particular number of quantum erasers. Thus, the ordered fringes of the intensity product of the cw light are the direct proof of the coherently excited superresolution. The quantum feature of superresolution and its interpretation sheds light on new quantum sensing metrology to overcome limited N00N states. The resulting phase bases of the ordered intensity products for superresolution may open the new wave of quantum technologies compatible with coherence optics-based conventional sensing metrologies.

**Methods**

The polarizing beam splitter (PBS) of the Michelson interferometer in Fig. 1 provides random polarization bases of the input photon (light) via a 22.5-degree rotated half-wave plate [8], where the role of the polarizers in the measurement blocks (see the Inset) is for the projection measurement of the quantum erasers [9]. By the PBS, polarization bases of the light are correlated to the paths. The vertical-basis photons are used as 'reference,' while the horizontal-basis photons are used as 'signal' to detect an unknown object 'target.' For the full collection of reference and signal photons into the output path, a quarter-wave plate is inserted in each path. For the intensity product, the output path ($E_A$) is divided into M sub-paths, where each path corresponds to each quantum eraser. For the phase control of the quantum eraser set in each block (see the dotted box in the Inset), a quarter-wave plate (QWP) is inserted, where the reference is block B without QWP. As experimentally demonstrated superresolution for N=4 [22], the block A is $\pm\pi/2$ phase shifted from the reference for the first-order intensity correlations. The maximum number N of light division is up to the photon numbers, where N is practically infinite because $10^{15}$ photons are in a 1 mW HeNe laser. For the quantum eraser, a pulsed laser scheme can be applied to distinguish different time-bin pulses for the intensity correlation measurements, where the physical distance between divided light (polarizers) through the PBS of the Michelson interferometer can be easily set to be beyond the light corn, satisfying the violation of the cause-effect relation.

**References**
1. Gerry, C. C. & Knight, P. L. Introductory Quantum Optics. (Cambridge Univ., Cambridge 2005). Ch. 6.
2. Boyd, R. W. Nonlinear Optics, Third Edition. New York (Academic Press, 2008) pp. 79–88.
3. Herzog, T. J., Kwiat, P. G., Weinfurter, H. & Zeilinger, A. Complementarity and the quantum eraser. Phys. Rev. Lett. **75**, 3034–3037 (1995).
4. Kim, T., Fiorentino, M. & Wong, F. N. C. Phase-stable source of polarization-entangled photons using a polarization Sagnec interferometer". Phys. Rev. A **73**, 012316 (2006).
5. Ham, B. S. The origin of anticorrelation for photon bunching on a beam splitter. Sci. Rep. **10**, 7309 (2020).
6. Ham, B. S. Coherently driven quantum features using a linear optics-based polarization-basis control. Sci. Rep. **13**, 12925 (2023).
7. Ham, B. S. The origin of Franson-type nonlocal correlation. arXiv:2112.10148 (2023).
8. Kim, S. and Ham, B. S. Observations of the delayed-choice quantum eraser using coherent photons. Sci. Rep. **13**, 9758 (2023).
9. Ham, B. S. Observations of the delayed-choice quantum eraser in a macroscopic system. arXiv:2205.14353v2 (2022).
10. Solano, E., Matos Filho, R. L. & Zagury, N. Deterministic Bell states and measurement of the motional state of two trapped ions. Phys. Rev. A **59**, R2539–R2543 (1999).
11. Ham, B. S. Phase-controlled coherent photons for the quantum correlations in a delayed-choice quantum eraser scheme. Sci. Rep. **14**, 1752 (2024).
12. Bohr, N. in *Quantum Theory and Measurement*, Wheeler, J.A. & Zurek, W.H. Eds. (Princeton Univ. Press, Princeton, NJ), pages 949, 1984.
13. Dirac, P. A. M. The principles of Quantum mechanics. 4th ed. (Oxford University Press, London), Ch. 1, p. 9 (1958).
14. Bohm, D. Quantum theory (Prentice-Hall, Inc. New Jersey, 1979). Ch. 6.
15. Grangier, P., Roger, G. and Aspect, A. Experimental evidence for a photon anticorrelation effect on a beam splitter: A new light on single-photon interferences. Europhys. Lett. **1**, 173-179 (1986).
16. Wheeler, J. A. In *Mathematical Foundations of Quantum Theory*, Marlow, A. R. Ed. (Academic Press, 1978), pp. 9–48.
17. Scully, M. O. & Drühl, K. Quantum eraser: A proposed photon correlation experiment concerning observation and "delayed choice" in quantum mechanics. Phys. Rev. A **25**, 2208–2213 (1982).
18. Jacques, V. et al. Experimental realization of Wheeler's delayed-choice Gedanken experiment. Science **315**, 966–978 (2007).
19. Kim, Y.-H., Yu, R., Kulik, S. P. and Shih, Y. Delayed "Choice" Quantum Eraser. *Phys. Rev. Lett*. **84**, 1-4 (2000).
20. Ham, B. S. A coherence interpretation of nonlocal realism in the delayed-choice quantum eraser. arXiv:2302.13474v4 (2023).
21. Sun, F. W., Liu, B. H., Gong, Y. X., Huang, Y. F., Ou, Z. Y. & Guo, G. C. Experimental demonstration of phase measurement precision beating standard quantum limit by projection measurement. EPL **82**, 24001

22. Kim, S. & Ham, B. S. Observations of super-resolution using phase-controlled coherent photons in a delayed-choice quantum eraser scheme. arXiv:2312.03343 (2023).
23. Jacobson, J., Gjörk, G., Chung, I. & Yamamato, Y. Photonic de Broglie waves. Phys. Rev. Lett. **74**, 4835–4838 (1995).
24. Walther, P. et al. Broglie wavelength of a non-local four-photon state. Nature **429**, 158–161 (2004).
25. Boto, A. N., Kok, P., Abrams, D. S., Braunstein, S. L., Williams, C. P. & Dowling, J. P. Quantum interferometric optical lithography: exploiting entanglement to beat the diffraction limit. Phys. Rev. Lett. **85** 2733 (2000).
26. Edamatsu, K., Shimizu, R. & Itoh, T. Measurement of the photonic de Broglie wavelength of entangled photon pairs generated by parametric down-conversion. Phys. Rev. Lett. **89**, 213601 (2002).
27. Dowling, J. P. Quantum optical metrology-the lowdown on high-N00N states. Contemp. Phys. **49**, 125-143 (2008).
28. Giovannetti, V., Lloyd, s. & Maccone, L. Quantum-enhanced measurement: beating the standard quantum limit. Science **306**, 1330-1336 (2004).
29. Nagata, T., Okamoto, R., O'Brian, J. L., Sasaki, K. & Takeuchi, S. Beating the standard quantum limit with four-entangled photons. Science **316**, 726-729 (2007).
30. Polino, E., Valeri, M., Spagnolo, N. & Sciarrino, F. Photonic quantum metrology. AVS Quantum Sci. **2**, 024703 (2020).
31. Aasi. J. et al. Enhanced sensitivity of the LIGO gravitational wave detector by using squeezed states of light. Nature Photon. **7**, 613-619 (2013).
32. Resch. K. J., Pregnell, K. L., Prevedel, R., Gilchrist, A., Pryde, G. J., O'Brien, J. L. & White, A. G. Time-reversed and super-resolving phase measurements. Phys. Rev. Lett. **98**, 223601 (2007).
33. Kothe, C., Björk, G. & Bourennane, M. Arbitrarily high super-resolving phase measurements at telecommunication wavelengths. Phys. Rev. A **81**, 063836 (2010).
34. Pedrotti, F. L., Pedrotti, L. M. & Pedrotti, L. S. Introduction to Optics, 3rd ed. (Pearson Education, Inc., New Jersey, 2004), Ch 14.
35. Pleinert, M.-O., von Zanthier, J. & Lutz, E. Many-particle interference to test Born's rule. Phys. Rev. Research **2**, 012051(R) (2020).
36. Ham, B. S. Deterministic control of photonic de Broglie waves using coherence optics. Sci. Rep. **10**, 12899 (2020).
37. Ham, B. S. Analysis of nonclassical features in a coupled macroscopic binary system. New J. Phys. **22**, 123043 (2020).
38. Boaron, A. *et al*. Simple 2.5 GHz time-bin quantum key distribution. Appl. Phys. Lett. **112**, 171108 (2018).
39. Kim, S. & Ham, B. S. Observations of super-resolution using phase-controlled coherent photons in a delayed-choice quantum eraser scheme. arXiv:2312.03343 (2023).
40. Torre, C. G. *Foundations of Wave Phenomena: Complete Version*. (DigitalCommons@USU, https://digitalcommons.usu.edu/foundation_wave/1, 2023), Ch. 4.

**Funding:** This research was supported by the MSIT (Ministry of Science and ICT), Korea, under the ITRC (Information Technology Research Center) support program (IITP 2024-2021-0-01810) supervised by the IITP (Institute for Information & Communications Technology Planning & Evaluation).

**Author contribution:** BSH solely wrote the paper.

**Competing Interests:** The author declares no competing interest.

**Data Availability Statement:** All data generated or analyzed during this study are included in this published article.